\documentclass[12pt]{article}
\newlength{\mytopmargin}
\newlength{\myleftmargin}
\usepackage{amsmath,amsthm,amssymb,amsbsy,epsfig,graphicx,color,multicol,subfigure}
\usepackage{array,calc}
\usepackage[enableskew]{youngtab}
\usepackage{rotating}
\usepackage{young,epic}
\usepackage{a4wide,bm}
\usepackage{url}
\usepackage{hyperref}

\usepackage{amsmath,amsfonts,amssymb}

\usepackage{graphicx}

\newcommand{\nc}{\newcommand}
\newtheorem{theorem}{Theorem}[section]

\newtheorem{example}[theorem]{Example}

\newtheorem{conj}{Conjecture}
\newtheorem{prop}{Proposition}
\nc{\bex}{\begin{example}}
\nc{\eex}{\end{example}}
\nc{\bea}{\begin{eqnarray}}
\nc{\eea}{\end{eqnarray}}
\nc{\ben}{\begin{eqnarray*}}
\nc{\een}{\end{eqnarray*}}
\nc{\bse}{\begin{subequations}}
\nc{\ese}{\end{subequations}}
\nc{\nn}{\nonumber}
\nc{\half}{\ensuremath{\frac{1}{2}}}
\nc{\Hom}{\operatorname{Hom}}
\nc{\End}{\operatorname{End}}
\nc{\vac}{|\textrm{vac}\rangle}
\nc{\tvac}{|\widetilde{\textrm{vac}}\rangle}
\nc{\dvac}{\langle\textrm{vac}}
\nc{\dtvac}{\langle\widetilde{\textrm{vac}}}
\nc{\id}{\mathbb{I}}
\nc{\Tr}{{\rm Tr}}
\nc{\tr}{{\rm Tr}}
\nc{\ree}[1]{\hbox{Re}({#1})}
\nc{\ws}{\;\;}
\newlength\celldim \newlength\fontheight \newlength\extraheight
\newcounter{sqcolumns}
\newcolumntype{S}{ @{}
>{\centering \rule[-0.5\extraheight]{0pt}{\fontheight + \extraheight}}
p{\celldim} @{} }
\newcolumntype{Z}{ @{} >{\centering} p{\celldim} @{} }
{\end{tabular}}

\nc{\ko}{\mbox{$k-1$}}
\nc{\kp}{\mbox{$k+1$}}
\nc{\kth}{\mbox{$k-3$}}
\nc{\kt}{\mbox{$k-2$}}
\nc{\nmo}{\mbox{$n-1$}}
\nc{\nmt}{\mbox{$n-2$}}
\nc{\nmth}{\mbox{$n-3$}}
\nc{\jo}{\mbox{$n-k-h+j+g+1$}}
\nc{\jt}{\mbox{$n-k-h+j+g+2$}}
\nc{\nkj}{\mbox{$n-k+j$}}
\nc{\nkgo}{\mbox{$n-h+g+1$}}
\nc{\nkg}{\mbox{$n-h+g$}}
\nc{\nkjo}{\mbox{$n-k+j+1$}}
\nc{\kho}{\mbox{$k+h-1$}}
\nc{\kht}{\mbox{$k+h-2$}}
\nc{\khth}{\mbox{$k+h-3$}}
\nc{\khf}{\mbox{$k+h-4$}}
\nc{\ho}{\mbox{$h-1$}}
\nc{\hp}{\mbox{$h+1$}}
\nc{\jkho}{\mbox{$j+k+h-1$}}
\nc{\jh}{\mbox{$n-k+j_1$}}
\nc{\jho}{\mbox{$n-k+j-1$}}
\nc{\jk}{\mbox{$n-h+1$}}
\nc{\nht}{\mbox{$n-h+2$}}
\nc{\jko}{\mbox{$n-h+g-1$}}
\nc{\nhgo}{\mbox{$n-h+g+1$}}
\nc{\jkt}{\mbox{$n-h+g-2$}}
\nc{\njhk}{\mbox{$n-k-h+j+g$}}
\nc{\mkh}{\mbox{Max$(k,h)$}}
\nc{\uo}{\mbox{$u+1$}}
\nc{\ut}{\mbox{$u+2$}}
\nc{\uth}{\mbox{$u+3$}}
\nc{\uno}{\mbox{$n+u-1$}}
\nc{\unt}{\mbox{$n+u-2$}}
\nc{\unkj}{\mbox{$n-k+u+j$}}
\nc{\unkjo}{\mbox{$n-k+u+j-1$}}
\nc{\nkog}{\mbox{$n-k-1+g$}}
\nc{\nhg}{\mbox{$n-k+g$}}
\nc{\nhjh}{\mbox{$n-h+1+j_h$}}
\nc{\nhjho}{\mbox{$n-k-h+2+j_{h-1}$}}
\nc{\nhjt}{\mbox{$n-k-1+j_{2}$}}
\nc{\nkhjo}{\mbox{$n-k+j_{1}+1$}}
\nc{\e}{\mbox{e}}
\nc{\ga}{\alpha}
\nc{\gb}{\beta}
\nc{\gd}{\delta}
\nc{\gep}{\varepsilon}
\nc{\gz}{\zeta}
\nc{\gt}{\theta}
\nc{\gk}{\kappa}
\nc{\gl}{\lambda}
\nc{\gp}{\phi}
\nc{\gs}{\sigma}
\nc{\go}{\omega}
\nc{\gn}{\nu}
\nc{\gr}{\rho}
\nc{\gm}{\mu}
\nc{\gu}{\upsilon}

\nc{\gou}{\underline{\go}}
\nc{\un}{\underline{n}}
\nc{\um}{\underline{m}}
\nc{\uw}{\underline{\go}}

\nc{\s}{\sigma}
\nc{\ep}{\varepsilon}
\nc{\z}{\zeta}
\nc{\g}{\gamma}
\nc{\zi}{\zeta^{-1}}

\nc{\gG}{\Gamma}
\nc{\gD}{\Delta}
\nc{\gT}{\Theta}
\nc{\gL}{\Lambda}
\nc{\gO}{\Omega}
\nc{\gP}{\Phi}

\nc{\cF}{\mathcal{F}}
\nc{\cP}{\mathcal{P}}
\nc{\cS}{\mathcal{S}}
\nc{\cN}{\mathcal{N}}
\nc{\cD}{\mathcal{D}}
\nc{\cH}{\mathcal{H}}
\nc{\cO}{\mathcal{O}}
\nc{\cT}{\mathcal{T}}
\nc{\cQ}{\mathcal{Q}}
\nc{\cW}{\mathcal{W}}
\nc{\cR}{\mathcal{R}}
\nc{\cC}{\mathcal{C}}

\nc{\pt}{\mathcal{P}\mathcal{T}}

\nc{\C}{\mathbb{C}}
\nc{\Q}{\mathbb{Q}}
\nc{\R}{\mathbb{R}}
\nc{\Z}{\mathbb{Z}}
\nc{\N}{\mathbb{N}}

\nc{\fg}{\mathfrak{g}}

\nc{\barx}{\bar{x}}
\nc{\bi}{\bar{i}}
\nc{\bj}{\bar{j}}
\nc{\bgr}{\bar{\rho}}
\nc{\bA}{\bar{\alpha}}
\nc{\bB}{\bar{\beta}}
\nc{\bC}{\bar{\gamma}}
\nc{\by}{\bar{y}}
\nc{\brv}{\overline{V}}
\nc{\brp}{\overline{P}}
\nc{\T}{\tilde{T}}
\nc{\tf}{\tilde{f}}
\nc{\te}{\tilde{e}}
\nc{\ts}{\tilde{s}}
\nc{\tgP}{\widetilde{\Phi}}
\nc{\tgPs}{\tilde{\Psi}}
\nc{\tgn}{\tilde{\nu}}
\nc{\tgl}{\tilde{\lambda}}
\nc{\tge}{\tilde{\eta}}
\nc{\txi}{\tilde{\xi}}
\nc{\tep}{\tilde{\epsilon}}
\nc{\tx}{\tilde{x}}

\nc{\cB}{\check{b}}
\nc{\cOm}{\check{\Omega}}

\nc{\goto}{\mapsto}

\begin{document}
%
%\begin{frontmatter}

\title{Asymptotics of spacing distributions 50 years later}
\author{Peter J. Forrester}
\date{}
\maketitle
\noindent
\thanks{\small Department of Mathematics and Statistics, 
The University of Melbourne,
Victoria 3010, Australia email:  p.forrester@ms.unimelb.edu.au }

%

%
%\author[ms]{Mark Sorrell}
%\ead{msorrell@unimelb.edu.au}
%

\begin{abstract}
\noindent
In 1962 Dyson used a physically based, macroscopic argument to deduce the first two terms of the large spacing  asymptotic expansion of the
gap probability for the bulk state of random matrix ensembles with symmetry parameter $\beta$. In the ensuing years,
the question of  asymptotic expansions of spacing distributions in random matrix theory has shown itself to have a rich mathematical
content. As well as presenting the main known formulas, we give an account of the mathematical methods used for their proofs, and
provide some new formulas. We also provide a high precision numerical computation of one of the spacing probabilities to illustrate the
accuracy of the corresponding asymptotics.
\end{abstract}

%\end{frontmatter}

\section{Introduction}
Random matrices were introduced in physics by Wigner in the 1950's \cite{Po65}.  Wigner's original hypothesis was that the statistical properties of energy levels of complex nuclei could be reproduced by considering an ensemble  of systems rather than a single system in which all interactions are completely described.  This allowed for an entirely mathematical approach where statistical properties of the spectrum of an ensemble of random matrices were considered.   But coming from physics, the aim was to use mathematics to compute experimentally measurable statistical quantities, and to compare against
the data. 

One viewpoint on a real spectrum from a random matrix is as a point process on the real line.  As such, perhaps the most natural statistical characterisation is that of the distribution of the eigenvalue spacing.  This choice of statistic becomes even more compelling when one considers that in many cases of interest, eigenvalue spectra can be `unfolded'.  This means that unlike many statistical mechanical systems, the density is not an independent control variable, but rather fixes the length scale only. Unfolding then is scaling the eigenvalues in the bulk of the spectrum so that the mean density is unity. 
It is indeed the bulk spacing distribution for the Gaussian orthogonal ensemble of real symmetric matrices --- albeit in an approximate form known as
the Wigner surmize (see e.g.~\cite{Me91}) --- which was compared against the empirical spacing distribution for the energy level of highly excited
nuclei (again see \cite{Me91}, and references therein).

Fixing length scales at the edge of the spectrum is, as a practical exercise, a more difficult task.  In addition to the bulk,
we will have interest in the soft and hard spectrum edges when the eigenvalue spectrum exhibits a square root profile and inverse square root profile respectively. 
To specify realisations of the bulk and edge regions of the eigenvalue spectrum, we recall (see e.g.~\cite{Fo10}) that the so-called classical random matrix ensembles have their eigenvalue probability density functions (PDFs) of the form
\bea\label{eigpdf}
\frac{1}{C}\prod_{l=1}^{N}g(\gl_l)\prod_{1\leq j <  k \leq N}|\gl_k-\gl_j|^\gb
\eea
with $\gb$ corresponding to the underlying global symmetry ($\gb=1,2$ or $4$ for invariance under orthogonal, unitary or symplectic unitary transformations respectively); $C$ denotes the normalization. This is extended to general $\gb>0$, giving the $\gb$-ensembles \cite{DE02} as specified by the eigenvalue
PDF (\ref{eigpdf}), 
to be denoted ME${}_{N,\beta}(g(\lambda))$.  In particular the choice $g(\gl)=\e^{-\gb\gl^2/2}$ defines the Gaussian $\gb$-ensemble and the choice $g(\gl)=\gl^{\gb a/2}\e^{-\gb\gl/2}$, $(\gl>0)$, defines the Laguerre $\gb$-ensemble.

The bulk state can be realized by scaling 
$
\gl_l \mapsto x_l/\sqrt{2N}
$
in the Gaussian $\gb$-ensemble.  The soft edge is realized by the scalings
$
\gl_l \mapsto \sqrt{2N}+\frac{x_l}{\sqrt{2}N^{1/6}}
$ and $ \gl_l \mapsto 4N+2\sqrt{2}x_l$
in the Gaussian and Laguerre $\gb$-ensembles respectively \cite{Fo93a}. Only the Laguerre $\gb$-ensemble has a hard edge, as it requires the eigenvalue density to be strictly zero on one side; it is realized by the scaling  
$
\gl_l \mapsto \frac{x_l}{4N} .
$
In all cases the limit $N \to \infty$ needs to be taken after the scaling. 
At an edge, the spacing between consecutive eigenvalues is not the natural observable.  Instead, it is most natural to measure the distribution of the largest, second largest etc.~eigenvalue (or smallest, second smallest etc.). It is well known, and easy to verify, that all these quantities can be
expressed in terms of the (conditional) gap probabilities $E_{\beta}^{(\cdot)}(n;J)$ for there being exactly $n$ eigenvalues in the interval $J$, for the
scaled state $(\cdot)=$ bulk, soft or hard indexed by $\beta$. In the case of the hard edge, the probability depends on the exponent $\beta a/2$ in the
Laguerre weight $\lambda^{\beta/2} e^{-\beta \lambda/2}$, so we write $E^{\rm hard}_\beta(n;J;a\beta/2)$.

Our interest in this review is on the asymptotic form of spacing distributions in the bulk, and of the distribution of large and small eigenvalues at the edge.  This is a topic which (in the bulk case) 
occupied the attention of Dyson in one of the pioneering papers on random matrix theory in the early 1960's \cite{Dy62e}, and is still being written on 
as we stand today some $50$ years later.  We are seeking to catalogue both the results, and the methods which underlie them, and also to contribute some
new formulas.
Section \ref{sec2} deals with results founded on Dyson's  heuristic physical hypothesis; these are in the form of conjectures. 
The various mathematical techniques which can both prove, and build on these asymptotic expressions, are covered in Section 3.
A numerical illustration of the accuracy of the asymptotic form is given in Section 4, as is a discussion of asymptotic results for  the gap probability
in the case that each eigenvalue is independently deleted with probability $(1 - \xi)$.\\

\section{Macroscopic heuristics}\label{sec2}
\subsection{Zero eigenvalues in the gap}
The eigenvalue PDF (\ref{eigpdf}) can be interpreted as the Boltzmann factor of a classical log-gas system interacting at inverse temperature $\gb$.  The particles repel via the logarithmic potential and are subject to a one body potential with Boltzmann factor $g(\gl) =\e^{-\gb V(\gl)}$.  This interpretation led Dyson 
\cite{Dy62e} 
to hypothesize an ansatz for the asymptotic form of the gap probability $E_{\beta}(0;(-\alpha, \alpha);{\rm C}\beta{\rm E}_N)$, where C$\beta$E${}_N$ denotes
Dyson's circular ensembles (see e.g.~\cite[Ch.~2]{Fo10})  of random unitary matrices (all eigenvalues are therefore on the unit circle; the interval
$(-\alpha, \alpha)$ refers to a sector of the circumference specified by its angles),
\bea\label{bdf}
E_{\beta}(0;(-\alpha, \alpha);{\rm C}\beta{\rm E}_N) \mathop{\sim}\limits_{N \to \infty} e^{-\beta \delta F}.
\eea
Here and below the symbol $\sim$ is used to denote that the RHS gives leading terms, up to some order to be further specified, of the asymptotic
expansion of the LHS.
In (\ref{bdf})
$\gd F$ is the energy cost of conditioning the equilibrium particle density so that $\gr _{(1)}(\theta)=0$ for $\theta \in (-\ga,\ga)$.  This energy cost consists of an electrostatic energy
\bea\label{V1}
V_1=-\frac{1}{2}\int_0^{2\pi}\int_0^{2\pi}\big(\gr_{(1)}(\theta_1)-N/2\pi\big)\big(\gr_{(1)}(\theta_2)-N/2\pi\big) \log|\e^{i\theta_1}-\e^{i\theta_2}|d\theta_1 d\theta_2
\eea
and an entropy term
\bea\label{V2}
V_2=\bigg( \frac{1}{\gb}-\frac{1}{2}\bigg) \int_0^{2\pi} \gr_{(1)}(\theta)\log\bigg( \frac{\gr_{(1)}(\theta)}{N/2\pi}  \bigg)
\eea
The density is chosen to minimize $V_1$ and then $V_1$ and $V_2$ evaluated, and we have
\begin{equation}\label{ft}
 \delta F = (V_1 + V_2).
 \end{equation}

\begin{prop}\label{densprop}(Dyson 1962)
With the requirements that $\gr_{(1)}(\theta)=0$ for $\theta \in (-\ga,\ga)$ and $\int_0^{2\pi}\gr_{(1)}(\theta)d\theta=N$, $V_1$ is minimized by
\bea
\gr_{(1)}(\theta)= \frac{N}{2 \pi} \frac{\sin (\theta/2)}{\sqrt{\sin^2 (\theta/2)-\sin^2(\ga/2)}}.
\eea
We then have
\bea \label{bv}
\gb V_1=-\frac{\gb}{2}N^2 \log \cos \big(\frac{\ga}{2}\big), \hspace{1cm} \gb V_2=\bigg( 1-\frac{\gb}{2} \bigg)N \log \bigg( \sec \big(\frac{\ga}{2}\big) +\tan \big(\frac{\ga}{2}\big)\bigg).
\eea
\end{prop}
We remark that explicit calculations in \cite{Dy62e} showed that requiring $\gr_{(1)}(\theta)$ to minimize $V_1+V_2$ (rather than $V_1$) results in a correction to $\gb V_2$ which for large $N$ is of order $\log(N\ga)$, indicating that the asymptotic expansion (\ref{bdf}) will not correctly give terms of this order.

Substituting (\ref{bv})  in (\ref{ft}), and substituting the result in (\ref{bdf})
gives a large deviation formula, telling us (as a conjecture) 
the probability of there being no eigenvalues in the interval $(-\alpha, \alpha)$. This probability decays as a Gaussian in $N$.
An $O(1)$ expression should result from choosing the excluded interval as $(-\pi s/N,\pi s/N)$, as then there are $O(1)$ eigenvalues in
the gap. Replacing $\ga$ by $\pi\ga/2$ in (\ref{bv}), then taking $N \to \infty$  (this is a double scaling limit) gives  the prediction
\begin{equation}
\lim_{N\to\infty}E_{\gb}(0;(-\pi s/N,\pi s/N);{\rm C}\beta{\rm E}_N)  
\sim  \e^{-\gb(\pi s)^2/16+(\gb/2-1)\pi s/2}.
\end{equation}
Dyson was well aware that the $\sim$ symbol  should be interpreted as agreeing in the large $s$ asymptotic expansion 
to the order given. But the LHS is the definition of $E_\beta^{\rm bulk}(0;(-s/2,s/2))$, thus providing the following conjecture.

\begin{conj} (Dyson 1962)
We have
\begin{equation}\label{D1}
E_\beta^{\rm bulk}(0;(0,s)) \mathop{\sim}\limits_{s \to \infty}  \e^{-\gb(\pi s)^2/16+(\gb/2-1)\pi s/2}.
\end{equation}
\end{conj}
As remarked above, Dyson  \cite{Dy62e} carried through the details of the minimization of $V_1+V_2$, resulting in a logarithmic correction to
the exponent of the RHS of (\ref{D1}): $((1-\beta/2)^2/(2 \beta)) \log s$. However, this was later put in doubt by des Cloizeaux and Mehta \cite{JMC72} who 
using a method based on eigenvalues (see Section \ref{SEig} below) obtained $-1/8$, $-1/4$ and $-1/8$ for the prefactor of $\log s$ for $\beta = $ 1,
2 and 4 respectively. In 1976 \cite{Dy76}, Dyson himself used inverse scattering methods applied to the Fredholm determinant form of
$E_1^{\rm bulk}(0;(0,s)) $ (see Section \ref{determ}) to also give the prediction $-1/8$ for the prefactor in the case $\beta = 1$. In fact the correct
extension of (\ref{D1}) for general $\beta$, as proved for the Gaussian $\beta$ ensemble, is \cite{VV08}
\begin{equation}\label{D2}
E_\beta^{\rm bulk}(0;(0,s))   \mathop{\sim}\limits_{s \to \infty}   \exp \Big ({-\gb(\pi s^2)/16+(\gb/2-1)\pi s/2 + (1/4)(\beta/2 + 2/\beta - 3) \log s + O(1)}
\Big ).
\end{equation}
Its derivation will be reviewed in Section \ref{SDE}.

The ansatz (\ref{bdf}) was applied to the gap probability at the hard edge of the Laguerre ensemble by Chen and Manning in 1994 \cite{CM94}.  They 
considered the probability of there being no eigenvalues in an interval $(0,t)$.  

\begin{prop}(Chen and Manning 1994)
For the Laguerre ensemble specified by (\ref{eigpdf})  with $g(\lambda) = \lambda^a e^{-\lambda}$, with the eigenvalues constrained to the
interval $(t,b)$, with $t > 0$ given, the 
minimizing solution for the level density $\gr_{(1)}(x)$ is 
\bea\label{cm1}
\gr_{(1)}(x) =\frac{1}{\pi\gb}\sqrt{\frac{b-x}{x-t}}\bigg( 1-\frac{a}{x}\sqrt{\frac{t}{b}}\bigg).
\eea
Normalization of the density requires that $b$ is related to $N$ by
\begin{equation}
N = {b - t \over 2 \beta} + {a \over \beta} \Big ( \sqrt{t \over b} - 1 \Big ).
\end{equation}
\end{prop}
Using (\ref{cm1}) appropriate analogues of (\ref{V1}) and
(\ref{V2}) were computed (see also \cite{CM96}), thus giving a prediction for the large $N$ form of
$E_\beta(0;(0,t);{\rm ME}_{N,\beta}(\lambda^\alpha e^{-\lambda}))$. This is exponentially small in $N$. But with
$t = s/(4N)$, the number of eigenvalues in $(0,t)$ will be $O(1)$. With the resulting expression interpreted as the 
large $s$ asymptotic form of $E_\beta^{\rm hard}(0;(0,s);a)$ ($s$ must be scaled $s \mapsto (\beta/2)^2 s$ to account for
the latter being defined as the large $N$ limit of $E_\beta(0;(0,s/(4N));{\rm ME}_{N,\beta}(\lambda^a e^{-\beta \lambda/2}))$),
the following conjecture was obtained.

\begin{conj} (Chen and Manning 1994)
We have
\begin{equation}\label{CM1}
E_{\gb}^{\rm hard}(0;(0,s);a) \mathop{\sim}\limits_{s \to \infty}
 \exp \Big ( {-\frac{\gb s}{8}+  a \sqrt{s}-\frac{a^2}{2\gb}\log s +\big(1-\frac{\gb}{2}\big)\frac{a}{2\gb}\log s} \Big ).
\end{equation}
\end{conj}
Historically, (\ref{CM1}) had already been proved for $a \in \mathbb Z_{\ge 0}$ and $2/\beta \in  \mathbb Z_{>0}$ in
\cite{Fo94b} before the work of \cite{CM94}. Moreover, the work \cite{Fo94b}, which was based on $a$-dimensional integral
forms for $E_{\gb}^{\mbox{\scriptsize{hard}}}(0;(0,s);a)$, gave the explicit form of the constant term in the extension of
(\ref{CM1}) to next order (see Section \ref{SH}).

The first application of the log-gas ansatz (\ref{bdf}) at the soft edge was due to
Dean and Majumdar \cite{DM06,DM08}.

\begin{prop}(Dean and Majumdar 2006)
Consider the Gaussian $\beta$-ensemble ME${}_{\beta,N}(e^{- \beta N x^2})$.  Suppose the eigenvalues are
confined to the interval $(-b, t)$ where $t < 1$ and $b> 0$ is determined by charge neutrality. The corresponding density is given by
$$
\rho_{(1)}(x) = {2N \over \pi} \Big ( {l - t + x \over t - x} \Big )^{1/2} \Big ( {l \over 2} - x \Big ),
$$
where $l := b + t = {2 \over 3}(t + \sqrt{t^2 + 3})$.
\end{prop}
Only the corresponding form of $V_1$ was computed, and this gave the large deviation formula
\begin{align}
& E_\beta(0;(t,\infty);{\rm ME}_{\beta,N}(e^{- \beta N x^2})) \nonumber \\
& \quad \mathop{\sim}\limits_{N \to \infty}
\exp \Big ( - \beta N^2\Big (
{2 t^2 \over 3} - {t^4 \over 27} - {5 \over 18} t \sqrt{3 + t^2} -
{1 \over 27} t^3 \sqrt{3 + t^2} - {1 \over 2} \log {t + \sqrt{t^2 + 3} \over 3}
\Big )\Big ),
\end{align}
and correspondingly, upon the appropriate soft edge scaling $\sqrt{2N}(1 - t) = -s/(\sqrt{2} N^{1/6})$, the asymptotic formula
\begin{equation}\label{CM1v}
E_\beta^{\rm soft}(0;(s,\infty)) \mathop{\sim}\limits_{s \to - \infty} e^{-\beta |s|^3/24}.
\end{equation}
This latter prediction was already implied by earlier work \cite{Fo93a}, \cite{TW94a}.

\subsection{Loop equations}
In 2011, Borot, Eynard, Majumdar and Nadal \cite{BEMN10} gave an 
alternative heuristic formalism to the Dyson log-gas ansatz, for purposes of computing the soft edge gap probability.
This in based on the so-called loop equations associated with the large $N$ form of the multiple integral definition of the latter.
The approach allows for the Dyson ansatz (\ref{bdf}) to be extended to include higher order terms; in practice  two new terms
are computed --- one is termed the Polyakov anomaly, and the following result is obtained.

\begin{conj} (Borot, Eynard, Majumdar and Nadal (2011))\label{CB}
We have
\begin{eqnarray}\label{mak}
\lefteqn{
E_\beta^{\rm soft}(0;(s,\infty)) \mathop{\sim}\limits_{s \to - \infty}  \exp \Big ( -\beta{ |s|^3 \over 24} +} \nonumber \\
&& 
{\sqrt{2} (\beta/2 - 1) \over 3} |s|^{3/2} + {(\beta/2) + (2/\beta) - 3 \over 8} \log |s| + \log \tau_\beta^{\rm soft} + O(|s|^{-3/2}) \Big ),
\end{eqnarray}
where
\begin{equation}\label{mak1}
 \log \tau_\beta^{\rm soft} = \Big ( {17 \over 8} - {25 \over 24} ((\beta/2) + (2/\beta)) \Big ) \log 2 -{  \log(2 \pi) \over 2} - {\log (\beta/2) \over 2} + \kappa_{\beta/2}
 \end{equation}
 with $\kappa_\beta$ the constant term in the large $N$ expansion of $F(N+1):=\sum_{j=1}^N \log \Gamma (1 + j \beta/2)$. (Note that in
 \cite{BEMN10} what we call $\beta/2$ is written as $\beta$.)
 \end{conj}

In  \cite{BEMN10}, for $\beta$ rational, $\kappa_\beta$ was evaluated in terms of the Barnes $G$-function, while for general $\beta > 0$ it
was shown
$$
\kappa_{\beta/2} = {\log (2 \pi) \over 4} + {\beta \over 2} \Big ( {1 \over 12} - \zeta'(-1) \Big ) + {\gamma \over 6 \beta} +
\int_0^\infty {1 \over e^{\beta s /2} - 1} \Big ( {s \over e^s - 1} - 1 + {s \over 2} - {s^2 \over 12} \Big ) \, ds,
$$
where $\gamma$ denotes Euler's constant. In fact $\kappa_{\beta/2}$ can be expressed in terms of the so-called Stirling modular form
$\rho_2(1,\tau)$, which from a computational viewpoint can be defined by the infinite product  \cite{Sh80}
$$
\rho_2(1,\tau) = (2 \pi )^{3/4} \tau^{-1/4 + (\tau + 1/\tau)/12} e^{P(\tau)} \prod_{n=1}^\infty {e^{Q(n\tau)} \over \Gamma(1 + n \tau)},
$$
where
$$
P(\tau) = - {\gamma \over 12 \tau} - {\tau \over 12} + \tau \zeta'(-1), \quad
Q(x) = \Big ( {1 \over 2} + x \Big ) \log x - x + \log \sqrt{2 \pi} + {1 \over 12 x}.
$$
The quantity $\rho_2(1,\tau)$ is fundamental to the theory of the Barnes double gamma function $\Gamma_2(z;1,\tau)$ \cite{Ba04}, the latter
being related to the usual gamma function through the two functional equations
\begin{equation}\label{r}
{1 \over \Gamma_2(z+1;1,\tau)} = {\tau^{z/\tau - 1/2} \over \sqrt{2 \pi}}
{\Gamma(z/\tau) \over \Gamma_2(z;1,\tau)}, \quad
{1 \over \Gamma_2(z+\tau;1,\tau)} = {1 \over \sqrt{2 \pi}} {\Gamma(z) \over  \Gamma_2(z;1,\tau)},
\end{equation}
and furthermore is normalized by requiring $\lim_{z \to 0} z \Gamma_2(z;1,\tau) = 1$.

\begin{prop}
Let $\tau = 2/\beta$ and specify $F(N+1)$ and $\kappa_{\beta/2}$ as in Conjecture \ref{CB}. We have
\begin{align}
F(N+1) & = (2 \pi)^{N/2} \tau^{-(N^2 - N(1 - \tau))/2 \tau} {\Gamma(N) \Gamma(1 + N/\tau) \over \Gamma_2(N;1,\tau)}, \label{T3} \\
\kappa_{1/\tau} & = - {1 \over 2} \log \tau + \log 2 \pi - \log \rho_2(1,\tau), \label{T4}
\end{align}
with the latter equation substituted into (\ref{mak1}) giving
\begin{equation}\label{T5}
 \log \tau_\beta^{\rm soft} = \Big ( {17 \over 8} - {25 \over 24} ((\beta/2) + (2/\beta) \Big ) \log 2 +{  \log(2 \pi) \over 2}  -
 \log \rho_2(1,2/\beta).
 \end{equation}
 \end{prop}
 
 The equation (\ref{T3}) has appeared in the recent work \cite{BMS11}; it follows immediately by characterizing $F(N+1)$ as
 a first order recurrence, and using (\ref{r}). The formula for $\kappa_{1/\tau}$ then follows by extracting the term independent of
 $N$ in the corresponding asymptotic expansion. Here one uses the fact that for $\log \Gamma_2(N;1,\tau)$ this is
 $\log \rho_2(1,\tau)$ \cite{QHS93}. A consequence of (\ref{T5}) is that
 \begin{equation}\label{tt}
 \log {\tau_{\beta/2}^{\rm soft} \over \tau_{2/\beta}^{\rm soft}} = - \log {\rho_2(1,2/\beta) \over \rho_2(1,\beta/2)} =
  {(\beta/2) + (2/\beta) - 3 \over 8} \log \Big ( {\beta \over 2} \Big )^{2/3} - {1 \over 2} \log {\beta \over 2},
  \end{equation}
  where the final equality follows from the inversion formula for the Stirling modular form \cite[Prop.~7(iv)]{KO98}.
 Using this in (\ref{mak}) gives that (cf.~\cite[Eq.~(6.2)]{BEMN10})
 \begin{equation}
 E_\beta^{\rm soft}(0;(s,\infty)) \mathop{\sim}\limits_{s \to - \infty}\Big ( {2 \over \beta} \Big )^{1/2}
  \widetilde{ E_{4/\beta}^{\rm soft}}(0;( ( {\beta \over 2}  )^{2/3} s,\infty)),
  \end{equation}
where $  \widetilde{ E_{\beta}^{\rm soft}}$ refers to the RHS of (\ref{mak}) with $ |s|^{3/2} $ replaced by $- |s|^{3/2} $.

\subsection{Conditioning $n$ eigenvalues in the gap}\label{nbulk}

In 1995, Dyson \cite{Dy95}, and independently Fogler and Shklovskii \cite{FS95},
further developed the log-gas argument by the consideration of the setting that the gap $(-t,t)$ is required
to contain exactly $n$ eigenvalues, with $0 \ll n \ll t$. Moreover, a change of viewpoint was introduced: the log-gas
was taken to be infinite in extent, with the bulk state characterized by a uniform density, normalized
to  unity. The $n$ eigenvalues are modelled as a continuous  conductive fluid occupying the interval
$(-b,b) \subset (-t,t)$. The electrostatic potential in this region must therefore be equal to a constant $-v$ say,
$v > 0$, with the potential in the other conducting region $\mathbb R \backslash (-t,t)$ taken to be zero.
The explicit form of the density was determined, and this substituted in the appropriate modification
of (\ref{V1}) and (\ref{V2}) gave after some calculation the simple results
\begin{equation}\label{Vv}
V_1 = - {n v \over 2} + {\pi^2 \over 4} (t^2 - b^2), \qquad V_2 = v.
\end{equation}
The end point $b$ is determined by $n$ via a certain elliptic integral, and similarly $v$ in terms of an elliptic integral
of modulus $b/t$. Expansion of these quantities for $t \to \infty$, and substitution in (\ref{bdf}) provides a generalization of
Conjecture \ref{D2}).

\begin{conj} \label{C4} (Dyson \cite{Dy95},   Fogler and Shklovskii \cite{FS95} (1995))
For $0 \ll n \ll s$ we have
\begin{multline}\label{Eb3}
\log E_{\beta}^{\rm bulk} (n; (0,s)) \mathop{\sim}\limits_{s \to \infty} 
  -\beta \frac{(\pi  s)^2} {16} + \left(\beta n +\frac{\beta} {2}- 1\right) \frac{\pi s}{2}\\
  +\left \{ \frac{n}{2} \left( 1- \frac{\beta}{2} -\frac{\beta n}{2}  \right )  +  \frac{1}{4}\Big (\frac{\beta}{2} + \frac{2}{\beta} - 3 \Big ) \right\}  \log  s
\end{multline}
(here we have added the $n=0$ contribution to the term $\log s$ as implied by (\ref{D2}) --- we then expect (\ref{Eb3}) to hold for
$0 \le n \ll s$; this is not a consequence of the calculations in
\cite{Dy95},  \cite{FS95}).
\end{conj}

Only very recently has this infinite log-gas formalism been applied to predict the asymptotic forms of the conditioned
gap probabilities at the hard and soft edges \cite{FW11}. Since the system is (semi-) infinite, this relies on characterizing
these edges in terms of the respective background densities: $\sqrt{x}/\pi$ for the soft edge, and $1/(2 \pi \sqrt{x})$ for the hard edge.
In both cases the coordinates are chosen so that the edge occurs at $x=0$. It was found in \cite{FW11} that applying
the ansatz (\ref{bdf}) with $\delta F$ given by (\ref{ft}) 
in this setting to the $n=0$ case gave results inconsistent with both (\ref{CM1}) and its soft edge
analogue in the second order term. Thus the ansatz (\ref{bdf}) with $\delta F$ given by (\ref{ft})  is incorrect in the infinite
log-gas formalism applied to the hard and soft edges. On the other hand, it was observed that replacing $V_2$ by the potential
drop $v$ in going from the region containing the infinite mobile log-charges, to the region containing the $n$ charges ---
which according to (\ref{Vv}) is an identity for the bulk --- restores the correct value for these terms. Making
this replacement for general $n$ then gives the following predictions.

\begin{conj} (Forrester and Witte \cite{FW11})
We have, for $0  \ll n \ll |s|$ (or more strongly $0 \le n \ll |s|$), 
\begin{multline}\label{22a}
\log E_\beta^{\rm hard}(n;(0,s);\beta a/2) \mathop{\sim}\limits_{s \to \infty}
- \beta \bigg\{ \frac{s}{8} - \sqrt{s} \left(n + \frac{a}{2} \right) \\
+ \Big[ \frac{n^2}{2} + \frac{na}{2}+ \frac{a(a-1)}{4} + \frac{a}{2\beta} \Big] \log s^{1/2} \bigg\},
\end{multline}
and
\begin{multline}\label{31h}
\log E_\beta^{\rm soft}(n;(s,\infty)) \mathop{\sim}\limits_{s \to - \infty } 
 - \frac{\beta|s|^3}{24} + \frac{2 \sqrt{2}}{3} |s|^{3/2} \left(\beta n + \frac{\beta}{2} -1\right)
 \\
 + \left[ \frac{\beta}{2} n^2 + \left( \frac{\beta}{2} - 1 \right) n + \frac{1}{6}  \left( 1 - \frac{2}{\beta} \left(1 - \frac{\beta}{2} \right)^2 \right) \right]\log |s|^{-3/4}.
\end{multline}
(As for (\ref{Eb3}), the results coming from the log-gas calculation have, in the case of the logarithmic term, been
supplemented by knowledge of the asymptotic expansion at that order for $n=0$.)
\end{conj}

We remark that a check on (\ref{Eb3}), (\ref{22a}) and (\ref{31h}) is that they obey certain asymptotic functional equations, implied
by exact functional equations for spacing distributions obtained in \cite{Fo09}. For example, at the hard edge one requires
$$
\label{23x} E_{\beta}^{ \rm hard} (n; (0,s/ \tilde{s}_{\beta});\beta a/2) \mathop{\mathop{\sim}\limits_{s\to \infty}}\limits_{n\ll t} 
 E_{4/\beta}^{ \rm hard} \left( \tfrac{1}{2}\beta(n+1)-1; \left( 0,s/\tilde{s}_{4/\beta} \right);a - 2 + 4/\beta \right),
$$
 where $\tilde{s}_{\beta}$ is an arbitrary length scale that satisfies $\tilde{s}_{4/ \beta} (\beta/2)^2 = \tilde{s}_{\beta}$. This is indeed a property
 of (\ref{22a}).
 
 Precise asymptotic statements can also be made concerning the asymptotic form of $E_\beta^{(\cdot)}(n;J)$, for $|J| \to \infty$ and
 $n \approx \langle n_J \rangle$, where $n_J$ ($ \langle n_J \rangle$) denotes the  number (expected number) of particles in $J$ for the unconstrained
 system. Thus macroscopic heuristics applied to this linear statistic (see e.g.~\cite[\S 14.5.1]{Fo11}) predict that $(n_J -  \langle n_J \rangle)/\sqrt{{\rm Var} \,
 n_J}$ has a Gaussian distribution with zero mean and unit variance, and so suggesting the following result.
 
 \begin{conj}
 For $n \approx  \langle n_J \rangle$,
 \begin{equation}\label{n0}
 E_\beta^{(\cdot)}(n;J) \mathop{\sim}\limits_{|J| \to \infty}
 {1 \over (2 \pi {\rm Var} \, n_J )^{1/2} } e^{- (n -   \langle n_J \rangle)^2 /2 {\rm Var} \, n_J}.
 \end{equation}
 Moreover, for $(\cdot) = $ bulk, soft and hard we have
 \begin{equation}\label{n1}
   \langle n_{(0,s)} \rangle \mathop{\sim}\limits_{s \to \infty} s, \quad
  \langle n_{(s,\infty)} \rangle \mathop{\sim}\limits_{s \to -\infty}   {2 (-s)^{3/2} \over 3 \pi}, \quad
  \langle n_{(0,s)} \rangle \mathop{\sim}\limits_{s \to \infty}  {s^{1/2} \over \pi}
  \end{equation}
  and
 \begin{equation}\label{n2}  
 {\rm Var} \, n_{(0,s)}  \mathop{\sim}\limits_{s \to \infty} {2 \over \pi^2 \beta} \log s, \quad
 {\rm Var} \, n_{(s,\infty)}  \mathop{\sim}\limits_{s \to -\infty} {1 \over \pi^2 \beta} \log |s|^{3/2}, \quad
  {\rm Var} \, n_{(0,s)}  \mathop{\sim}\limits_{s \to \infty} {1 \over \pi^2 \beta} \log s^{1/2}. 
   \end{equation} 
 \end{conj}
The results (\ref{n1}) are immediate consequences of the corresponding asymptotic density profiles (recall the second
sentence below Conjecture \ref{C4}), while (\ref{n2}) can be derived heuristically from knowledge of the asymptotic form of the
two-point correlation function (see \cite[paragraph below (14.87)]{Fo10}). In the case of $(\cdot)=$ bulk, (\ref{n0}), with the corresponding
vaues of $   \langle n_{(0,s)} \rangle $ and $ {\rm Var} \, n_J $ as implied by (\ref{n1}) and (\ref{n2}), was derived in the context of the infinite
log-gas formalism by Dyson \cite{Dy95} and by Fogler and Shklovskii \cite{FS95}.

\section{Rigorous methods}
\subsection{Toeplitz/Hankel asymptotics}\label{determ}
% Krasovsky,Deift    \\ \\
It is a fundamental result in random matrix theory (see e.g.~\cite[\S 9.1]{Fo10}) that in the scaled limit $(\cdot)$ equal to bulk, hard or soft, and $\beta = 2$
 the probability of there being no eigenvalues in an interval $J$, may be written in terms of a determinant of a Fredholm integral operator
 \ben
 E^{(\cdot)}_{2}(0;J)=\det (1- K_J^{(\cdot)}),
 \een
where $K_J^{(\cdot)})$ is the integral operator on the interval $J$ with well known sine, Bessel and Airy kernels (see e.g.~\cite{Fo10} for the precise
definitions). This is related to the fact that for $\beta = 2$ the gap probabilities can be written in terms of either Toeplitz or Hankel determinants.
For example, the Toeplitz determinant of a function $f(\theta)$, integrable over the unit circle, is defined as
\bea
D_n(f):=\det\bigg( \frac{1}{2\pi}\int_{0}^{2\pi}\e^{-i(j-k)\theta}f(\theta)d\theta \bigg)_{j,k=0}^{n-1},
\eea
and one has the well known formula
$$
E_2(0;(-\alpha,\alpha);{\rm CUE}_N) = D_N(f_\alpha), \qquad f_\alpha = \left \{ 
\begin{array}{ll} 1, & \theta \in (\alpha, 2 \pi - \alpha) \\
0, & {\rm otherwise} \end{array} \right.
$$
In particular $\lim_{n\to\infty}D_n(f_{2s/n})=\det (\id -  K_s^{\mbox{\scriptsize{bulk}}})$, allowing for a strategy whereby the 
$s \to\infty$ behaviour can be extracted from the asymptotics of the Toeplitz determinant. On the other hand the Toeplitz determinant has a representation in terms of quantities associated with orthonormal polynomials $\phi_k(z) = \chi_k  z^k+\dots$ with weight $f(\theta)$  on the unit circle;
explicitly $D_n(f)=\prod_{k=0}^{n-1}\chi_k^{-2}$. Krasovsky \cite{Kr04} used a Riemann-Hilbert formulation to compute the large $n$ form of
$\frac{d}{d\mu} \ln D_n(f_\mu)$, uniformly in $\mu$, providing both a proof and refinement of (\ref{D1}) in the case $\beta = 2$.

\begin{theorem}(Krasovksy 2004, Ehrhardt 2006)
We have
\bea\label{3.23}
\log E_2^{\rm bulk}(0;(0,s)) =-\frac{(\pi s)^2}{8}-\frac{1}{4} \log (\pi s/2) + \frac{1}{12} \log 2 +3\zeta'(-1)+O\bigg(  \frac{1}{s} \bigg)
\eea
where $\zeta (z)$ is the Riemann zeta-function.
\end{theorem}
We remark that up to the constant term this result, deduced by Dyson \cite{Dy76} using a scaling argument from 
known  Toeplitz determinant asymptotics, was first rigorously proved by  Deift,  Its and Zhou \cite{DIZ96}; also the proof of Ehrhardt 
\cite{Eh06}
is operator theoretic, and does not make use of orthogonal polynomials.

Analogous strategies can be used to analyze the hard and soft edges for $\beta = 2$, giving the following results, proving and extending
(\ref{CM1}) and (\ref{CM1v}) respectively. 

\begin{theorem} (Deift, Its and Krasovsky 2008, and  Deift, Krasovsky and Vasilevska 2010)
We have, for $s \to -\infty$ \cite{DIK08}
\begin{equation}\label{D1x}
\log E_2^{\rm soft}(0;(s,\infty)) = -\frac{|s|^3}{12}-\frac{1}{8}\log |s| +\frac{1}{24} \ln 2 +\zeta'(-1)+O(|s|^{-3/2})
\end{equation}
and for $s \to \infty$ \cite{DKV11}
\begin{equation}\label{D2x}
\log E_2^{\rm hard}(0;(0,s);a) = - {s \over 4} + a \sqrt{s} - {a^2  \over 4} \log s + \log \Big ( \frac{G(1 + a)}{(2 \pi)^{a/2}} \Big )+ O(s^{-1/2})
\end{equation}
where $G(x)$ denotes the Barnes $G$-function.
\end{theorem}
An alternative proof of (\ref{D1x}) has been given by Baik, Buckingham and DiFranco in 2008 \cite{BBD07},
using the Painlev\'e form of $ E_2^{\rm soft}(0;(s,\infty))$ \cite{TW94a}. This method carries over to the cases $\beta = 1$ and 4,
and in \cite{BBD07} the expansion (\ref{mak}) with
\begin{equation}\label{mak2}
\log \tau_1^{\rm soft} = - {11 \log 2 \over 48} + {\zeta'(-1) \over 2}, \qquad
\log \tau_4^{\rm soft} = - {37 \log 2 \over 48} + {\zeta'(-1) \over 2}
\end{equation}
was obtained. These confirm the values implied by (\ref{T5}).

With regards to  (\ref{D2x}), 
as noted above, for $a \in \mathbb Z_{\ge 0}$ it was
first proved by Forrester \cite{Fo94b}. More recently a proof of (\ref{D2x}) valid for $|a| < 1$ was given by Ehrhardt \cite{Eh10}.
Furthermore, let the next order (constant) term in the exponent of  (\ref{CM1}) be included by adding $\log \tau_{a,\beta}^{\rm hard}$.
We read off from (\ref{D2x}) that $  \tau_{a,2}^{\rm hard} =
G(1+a)/(2\pi)^{a/2}$. For $a \in \mathbb Z_{\ge 0}$ a multiple integral form for $E_1^{\rm hard}$ \cite{FW01a}, and an identity  \cite{FR01} relating
$E_4^{\rm hard}$ to $E_2^{\rm hard}$ and $E_1^{\rm hard}$ for general $a > -1$ tells us that
\begin{equation}\label{3.26a}
\tau_{a,1} = 2^{-a(a+1/2)} {G(3/2) G(2a+2) \over G(a+3/2) G(a+2)}, \qquad
\tau_{a+1,4} =  2^{-a(a+1)/4-1} {\tau_{a,2} \over \tau_{(a-1)/2,1}}.
\end{equation}

In the case of bulk scaling, include a constant term by adding  $\log \tau^{\rm bulk}_\beta$ to the exponent of  (\ref{D2})
with $s$ replaced by $s/\pi$ (thus the bulk density is now $1/\pi$).
It follows from (\ref{3.23}) that $\log \tau_2^{\rm bulk} =  {1 \over 3} \log 2 + 3 \zeta'(-1)$.
And inter-relations between the bulk gap probability for $\beta = 1$ and 4 with $\beta=2$ quantities give that 
 \cite{BTW92},
\begin{equation}\label{3.26b}
 \tau_1^{\rm bulk} = 2^{5/12} e^{(3/2) \zeta'(-1)} , \qquad
 \tau_4^{\rm bulk} = 2^{-29/24} e^{(3/2) \zeta'(-1)} .
 \end{equation}
 We observe that (\ref{3.26b}) is consistent with a relation analogous to (\ref{T5}).
 
 \begin{conj}
 Let $\rho_2(1,\tau)$ denote the Stirling modular form. We have
 \begin{equation}\label{tt2}
  \log  \tau_{\beta/2}^{\rm bulk} = \Big (3  - {4 \over 3} (\beta/2 + 2/\beta)\Big ) \log 2 + 3 \Big ( {1 \over 2} \log 2 \pi - \log \rho_2(1,2/\beta) \Big ),
  \end{equation}
  and consequently
 \begin{align}\label{tt1a}
 \log {\tau_{\beta/2}^{\rm bulk} \over \tau_{2/\beta}^{\rm bulk}} & = - 3 \log {\rho_2(1,2/\beta) \over \rho_2(1,\beta/2)}, \nonumber \\
 E_\beta^{\rm bulk}(0;(0,s/\pi ))& \mathop{\sim}\limits_{s \to \infty} \Big ( {2 \over \beta} \Big )^{3/2}
  \widetilde{ E_{4/\beta}^{\rm bulk}}(0;( 0,{\beta \over 2} s/\pi)),
  \end{align}
where $  \widetilde{ E_{\beta}^{\rm bulk}}$ refers to the RHS of (\ref{D2}) with $s$ replaced by $- s$ in the second term.

 \end{conj}

\subsection{Stochastic differential equations}\label{SDE}
The Gaussian and Laguerre $\beta$-ensembles, defined as eigenvalue PDFs below (\ref{eigpdf}), admit realizations as real symmetric tridiagonal
matrices \cite{DE02}. In the scaled $N \to \infty$ limit, this in turn leads to explicit characterization of gap probabilities in terms of
stochastic differential equations. The first result of this type was done for the soft edge, by Ramirez, Rider and Valko \cite{RRV06}. With $N$ fixed,
it relies on expressing the number of eigenvalues greater than $\mu$ as the number of sign changes of the shooting vector for the
tridiagonal matrix. Similarly at the hard edge \cite{RR08}. In the bulk, the shooting eigenvector must be parametrized in terms of the
corresponding Pr\"ufer phase \cite{KS06, VV07}. The following results are obtained.

\begin{prop}
Let $b_t$ denote standard Brownian motion. At the soft edge, define a diffusion by the Ito process \cite{RRV06}
$$
dp(t) = {2 \over \sqrt{\beta}} db_t + (\lambda + t - p^2(t)) dt, \quad p(0) = \infty;
$$
at the hard edge with parameter $\beta(a+1)/2 - 1$ by \cite{RRZ11}
$$
dp(t) = db_t + \Big ( {\beta \over 4} (a + {1 \over 2}) - {\beta \over 2} \sqrt{\lambda} e^{- \beta t/8} \cosh p(t) \Big ) dt, \quad  \quad p(0) = \infty;
$$
and in the bulk by \cite{VV08}
$$
dp(t) = db_t + \Big ( {1 \over 2} \tanh p(t)  - {\beta \over 8} \lambda  e^{- \beta t/4} \cosh p(t) \Big ) dt, \quad  \quad p(0) = \infty.
$$
Let $J = (0,s/2 \pi)$ for $(\cdot)=$ bulk; $J = (0,s)$ for $(\cdot)=$ hard; $J=(s,\infty)$ for $(\cdot)=$ soft. We have
\begin{equation}\label{EP}
E_\beta^{(\cdot)}(0;J) = {\rm Pr}(p(t) > - \infty, \forall t \in \mathbb R^+ \cup \{ \infty \} ).
\end{equation}
\end{prop}

The utility of these characterizations for the purpose of asymptotics is that they allow, via the Cameron-Martin-Girsanov formula, (\ref{EP}) to be
rewritten as the expectation of a functional of a transformed stochastic process. In contrast to (\ref{EP}), this functional allows for
a systematic, rigorous $s \to \infty$ asymptotic analysis  resulting in a proof of (\ref{D2}) --- giving in the process the correct form
of the general $\beta > 0$, $\log s$ term, for the first time --- and a proof of (\ref{CM1}) for general $\beta > 0$ and $a>-1$.
At the soft edge only the leading asymptotic form (\ref{CM1v}) has been proved using this approach \cite{RRV06}.

For the large $N$ limit of the circular $\beta$-ensemble, the Pr\"ufer phase has been used to prove the analogue of the Gaussian fluctuation formula
(\ref{n0}), namely $E_\beta(n,(-\alpha,\alpha);{\rm C}\beta {\rm E}_N) \sim (1/(2 \pi {\rm Var} \, n_{(-\alpha,\alpha)}))
 \exp(-(n - N\alpha/\pi)^2/(2  {\rm Var} \, n_{(-\alpha,\alpha)}))$, where $ {\rm Var} \, n_{(-\alpha,\alpha)} \sim (1/(\pi^2 \beta)) \log N$ \cite{Ki08} .
\subsection{Fredholm determinant/eigenvalue forms for $E_\gb^{(\cdot)}(n,J)/E_\gb^{(\cdot)}(0,J)$}\label{SEig}
With $(\cdot)$ denoting bulk, soft or hard, let $E^{(\cdot)}_\gb(J;\xi)$ be the generating function for $\{ E_\gb^{(\cdot)}(n;J) \}$, so that 
\bea \label{dilute}
E^{(\cdot)}_\gb(J;\xi)=\sum_{n=0}^\infty (1-\xi)^n E_\gb^{(\cdot)}(n;J).
\eea
Generalising the Fredholm determinant expressions for $E^{(\cdot)}_\gb(0;J)$ from Section \ref{determ}, one has that for $\beta = 2$
\bea \label{dilute1}
E^{(\cdot)}_2(J;\xi)=\det (1-\xi K_J^{(\cdot)})=\prod_{l=0}^{\infty}(1-\xi \gl_l),
\eea
where $1>\gl_0>\gl_1>\gl_2>\dots > 0$ are the eigenvalues of $K^{(\cdot)}_{J}$.  Consequently 
\bea\label{EJ}
\frac{E^{(\cdot)}_2(n;J)}{E^{(\cdot)}_2(0;J)}=\sum_{0\leq j_1<\dots<j_n}\frac{\gl_{j_1}\dots\gl_{j_n}}{(1-\gl_{j_1})\dots(1-\gl_{j_n})}.
\eea
It has been known since the work of Gaudin \cite{Ga61} that associated with $K^{\mbox{\scriptsize{bulk}}}_{(0,s)}$ is a commuting differential operator.  Furthermore, the work of Fuchs \cite{Fu64} uses this, together with a WKB asymptotic analysis, to deduce the $s \rightarrow \infty$ asymptotic form of $\gl_j$ ($j$ fixed).  It was noted by Tracy and Widom \cite{TW93} that the latter implies the term with $(j_1,j_2,\dots,j_n)=(0,1,\dots,n-1)$ dominates as $t\rightarrow \infty$.  These authors carried out a similar analysis in the soft and hard edge cases \cite{TW94a,TW94b}, so arriving at the following result (stated as Prop. 9.6.6 in \cite{Fo10}). 
\begin{prop}
Let $G(x)$ denote the Barnes $G$-function.  For $n$ fixed
\begin{eqnarray}\label{eigexp}
\frac{E_2^{\rm bulk }(n;(0,s))}{E_2^{\rm bulk}(0;(0,s)) } &\underset{s \rightarrow \infty}{\sim}& G(n+1)\pi^{-n/2}2^{-n^2-n}(\pi s)^{-n^2/2}\e^{n\pi s}\nn \\
\frac{E_2^{\rm soft }(n;(0,s))}{E_2^{\rm soft}(0;(0,s)) } &\underset{s \rightarrow \infty}{\sim}&\nn \frac{G(n+1)}{\pi^{n/2}2^{(5n^2+n)/2}}(-s/2)^{-3n^2/4} { \rm exp} \Bigg( \frac{8n}{3}\bigg(-\frac{s}{2}\bigg)^{3/2}\Bigg) \\
\frac{E_2^{\rm hard }(n;(0,s))}{E_2^{\rm hard}(0;(0,s))}  &\underset{s \rightarrow \infty}{\sim}&\frac{G(a+n+1)G(n+1)}{G(a+1)}\pi^{-n}2^{-n(2n+2a+1)}s^{-n^2/2-an/2}\e^{2n\sqrt{s}}.
\end{eqnarray}
\end{prop}
In \cite[\S 9.6.2]{Fo10}, as $t\to \infty$,  $E_1^{\mbox{\scriptsize{bulk}}}(n;t)$ and  $E_4^{\mbox{\scriptsize{bulk}}}(n;t)$ are related to $E_2^{\mbox{\scriptsize{hard}}}(\cdot;\cdot)$ for particular choices of the parameters.  The asymptotics of the latter are known as noted in the above proposition, allowing us to extend the first result in (\ref{eigexp}) to $\gb=1$ and $4$ (\cite{Fo10}, eqns (9.100) and (9.102)).
\begin{prop}\label{x1}
For $n$ fixed and $\gb=1$ and $4$ we have
\bea\label{beta14}
\frac{E_\gb^{\rm bulk }(n;(0,s))}{E_\gb^{\rm bulk }(0;(0,s))}=c_{\gb,n} \frac{\e^{\gb n \pi s/2}}{(\pi s)^{\gb n^2/4+(\gb/2-1)n/2}}\Bigg( 1+O \bigg(\frac{1}{s} \bigg) \Bigg),
\eea
where
\ben
c_{1,n}&=& \frac{G(n/2+1/2)G(n/2+1)}{G(1/2)}\pi^{-n/2}2^{-n(n+1)/4}, \\
c_{4,n}&=&\frac{G(n+3/2)G(n+1)}{G(3/2)}\pi^{-n}2^{-2n(n+1)} . 
\een
\end{prop}
According to the first asymptotic formula in (\ref{eigexp}), (\ref{beta14}) is, for a specific $c_{2,n}$, valid too for $\gb=2$.  Furthermore the functional form \ref{beta14} for general $\gb>0$ coincides with the log-gas prediction (\ref{31h}), and thus validates the latter for $\gb=1,2$ and $4$, and furthermore extends it by the evaluation of $c_{\gb,n}$.  

We would like to extend Proposition \ref{x1} to the soft and hard edge cases.  For this, let $V_{\tilde{J}}^{(\cdot)}$ for $(\cdot )=$ soft, hard and $\tilde{J}=(0,\infty), (0,1)$ respectively denote the integral operators on $\tilde{J}$,  dependent on a parameter $s$, with kernels Ai$(x+y+s)$ and $\frac{\sqrt{s}}{2}J_a(\sqrt{sxy})$. 
Write
\ben
E_{\pm}^{(\cdot)}(\xi;J)=\det (\id \mp \sqrt{\xi}V_{\tilde{J}}^{(\cdot)})
\een
and define 
\ben
E_{\pm}^{(\cdot)}(n;J):=\frac{(-1)^n}{n!}\frac{\partial^n}{\partial \xi^n} E_{\pm}^{(\cdot)}(\xi;J)\bigg|_{\xi=1}.
\een
Results contained in \cite{Fo06c}, and further refined in \cite{Bo09}, tell us that for $s \to \infty$
\bea\label{E114}
E_1^{\mbox{\scriptsize{soft}}}(2k;(s,\infty))&=& E_+^{\mbox{\scriptsize{soft}}}(k;(s,\infty))+\dots \nn \\
E_1^{\mbox{\scriptsize{soft}}}(2k+1;(s,\infty))&=&\frac{1}{2}E_-^{\mbox{\scriptsize{soft}}}(;(s,\infty))+\dots \nn \\
E_4^{\mbox{\scriptsize{soft}}}(k;(s,\infty))&=& E_1^{\mbox{\scriptsize{soft}}}(2k+1;(2^{2/3}s,\infty))+\dots  
\eea
Here terms not written on the RHS are exponentially smaller (in $s$) than the given term. \\
To proceed further requires a property of the eigenvalues of $V_{\tilde{J}}^{(\cdot)}$ which although supported by
numerical computations, to our knowledge  is yet to be proven.
\begin{conj}\label{C1}
Let $\gu_j^{(\cdot)} (j=0,1,2,\dots)$ denote the eigenvalues of $V_{\tilde{J}}^{(\cdot)}$, ordered so that $|\gu_0^{(\cdot)}|<|\gu_1^{(\cdot)}|<|\gu_2^{(\cdot)}|<\dots$.  Then $\gu_{2j}^{(\cdot)}>0$ while $\gu_{2j+1}^{(\cdot)}<0$ for each $j=0,1,2,\dots$.
\end{conj}

It is well known, and easy to verify (see e.g.~\cite[\S 9.6.1]{Fo10})  that $(\gu^{(\cdot)}_j)^2=\gl_j^{(\cdot)}$, where $\{\gl_j^{(\cdot)}\}$ are the eigenvalues of $K_J^{(\cdot)}$.  This fact, together with Conjecture \ref{C1} and the analogue of (\ref{EJ})  relating to$E_\pm^{\mbox{\scriptsize{soft}}}(k;(s,\infty))$, tells us that for $s \to \infty$
\bea \label{E115}
\frac{E_1^{\mbox{\scriptsize{soft}}}(2k;(s,\infty))}{E_1^{\mbox{\scriptsize{soft}}}(0;(s,\infty))}&=&\frac{1}{(1-\gl_0^{\rm s})(1-\gl_2^{\rm s})\dots(1-\gl_{2k-2}^{\rm s})}+\dots \nn \\
\frac{E_1^{\mbox{\scriptsize{soft}}}(2k+1;(s,\infty))}{E_4^{\mbox{\scriptsize{soft}}}(0;(s/2^{2/3},\infty))}&=&\frac{1}{(1-\gl_1^{\rm s})(1-\gl_3^{\rm s})\dots(1-\gl_{2k-1}^{\rm s})}+\dots
\eea
Knowledge of the explicit asymptotic form of $\gl_j^s$ from \cite{TW94a}, together with the asymptotic form of $E_1^{\mbox{\scriptsize{soft}}}(0;(s,\infty))/E_4^{\mbox{\scriptsize{soft}}}(0;(s/2^{2/3},\infty))$ implied by (\ref{mak}) and (\ref{mak2})  then allows us to extend the second result of (\ref{eigexp}) to $\gb=1$ and $4$.\\
\begin{prop} \label{P6} (under the assumption of Conjecture \ref{C1}) \\
We have 
\bea
\frac{E_1^{\rm soft }(n;(s,\infty))}{E_1^{\rm hard }(0;(s,\infty) } &\underset{s \rightarrow -\infty}{\sim}& \frac{G(n/2+1/2)G(n/2+1)}{\pi^{n/2}G(1/2)}2^{-(5/8)n^2+(1/8)n}(- s)^{-(3/8)n^2+(3/8)n}  \nn \\
& & \times {\rm exp}\Bigg( \frac{4n}{3}\bigg( -\frac{s}{2}\bigg)^{3/2}  \Bigg)\nn \\
\frac{E_4^{\rm soft }(n;(s/2^{2/3},\infty))}{E_4^{\rm soft }(0;(s/2^{2/3},\infty)) } &\underset{s \rightarrow -\infty}{\sim}&\sqrt{2}\e^{-\frac{\sqrt{2}}{3}(-s)^{3/2}} 
\frac{E_1^{\rm soft}(2n+1;(s,\infty))}{E_1^{\rm soft }(0;(s,\infty)) }
\eea
\end{prop}

At the hard edge, formulas structurally identical to (\ref{E114}) hold \cite{Fo06c}, \cite{Bo09}, with the important
qualification that the additional label need to specify the hard edge gap probabilities is $(a-1)/2$ on the LHS of the first two equations,
and $a+1$ on the LHS of the third equation; on the RHS's it is  $a,a$ and $a-1$ respectively, and in the third
equation $s$ is scaled by $4$ instead of $2^{2/3}$. The analogue of  (\ref{E115}) then allows the analogue of Proposition \ref{P6} to be deduced.

\begin{prop} \label{P7} (under the assumption of Conjecture \ref{C1}) \\
We have
\bea
\frac{E_1^{\rm hard }(n;(0,s);(a-1)/2)}{E_1^{\rm hard }(0;(0,s);(a-1)/2) }&\underset{s \rightarrow \infty}{\sim}&
2^{-n(n-1+a)/2} (2 \pi)^{-n} \prod_{p=1}^2 {G((n+p)/2) G((n+p+a)/2) \over G(p/2) G((p+a)/2)}
 \nn \\
& & \times s^{-(n^2 + n (a - 1))/4}  e^{n\sqrt{s} } \nn \\
\frac{E_4^{\rm hard }(n;(0,s/4);a+1)}{E_4^{\rm hard }(0;(0,s/4);a+1) } &\underset{s \rightarrow \infty}{\sim}&
e^{- \sqrt{s}} s^{a/4} {2^{(a+1)/2} (2 \pi)^{1/2} \over \Gamma((a+1)/2)}
\frac{E_1^{\rm hard }(2n+1;(0,s);(a-1)/2)}{E_1^{\rm hard }(0;(0,s);(a-1)/2) }. \nn
\eea

\end{prop}

As a final point in this subsection, we remark that the Gaussian fluctuation formula (\ref{n0}) can be proved for $\beta = 2$, using only the determinantal structure (\ref{dilute1})  and the
fact that Var$\, n_J \to \infty$ \cite{CL95,So00}.

\subsection{Hard edge -- generalized hypergeometric functions}\label{SH}
In the case that $a \in \Z^+$ and general $\gb>0$, the hard edge gap probability $E^{\mbox{\scriptsize{hard}}}_{\gb}(0;(0,s);a)$ permits evaluation in terms of a generalized hypergeometric function based on Jack polynomials $P_\kappa^{(\alpha)}(z_1,\dots,z_N)$.  The latter are labelled by a partition $\kappa_1\geq \kappa_2 \geq \dots \geq \kappa_N\geq 0$ of non-negative integers, and depend on the parameter $\alpha$.  For $\alpha=1$ they are the Schur polynomials, while for $\alpha=2$ they are the zonal polynomials of mathematical statistics; their precise definition can be found in  e.g.~\cite[\S 12.6]{Fo10}.  Defining $C_\kappa^{(a)}(z_1,\dots,z_N)$ as proportional to $P_\kappa^{(\alpha)}(z_1,\dots,z_N)$ with a specific proportionality depending on $\alpha$ and $\kappa$  
\cite[eq.~(13.1)]{Fo10}, and the generalised Pochhammer symbol 
$[u]^{(a)}_\kappa$ \cite[eq.~(12.46)]{Fo10},
the generalized hypergeometric function ${}_p F_q^{(\alpha)}$ is specified by (see e.g.~\cite[\S 13.1]{Fo10})
\bea 
_p F_q^{(\alpha)}(a_1,\dots,a_p;b_1,\dots,b_q;x_1,\dots,x_m):=\sum_\kappa \frac{1}{|\kappa|!}\frac{[a_1]^{(\alpha)}_\kappa\dots [a_p]^{(\alpha)}_\kappa }{[b_1]^{(\alpha)}_\kappa
\dots [b_q]^{(\alpha)}_\kappa} 
C_\kappa^{(\alpha)}(x_1,\dots,x_m).
\eea
Like their classical counterpart, these exhibit the confluence property
$$
\lim_{a_p\rightarrow \infty}  {}_p F_q^{(\alpha)}(a_1,\dots,a_p;b_1,\dots,b_q;{ x_1 \over a_p},\dots, {x_m \over a_p})= {}_{p-1} F_q^{(\alpha)}(a_1,\dots,a_{p-1};b_1,\dots,b_q;x_1,\dots,x_m). 
$$
Using this in the case $p=q=1$, together with an integral expression for ${}_1 F_1$  \cite[\S 13.2.5]{Fo10}
we can readily express the conditional gap probability  $E^{\mbox{\scriptsize{hard}}}_{\gb}(n;(0,s);a)$ for $a,\gb\in \Z_{\geq 0}$ in terms of the generalized hypergeometric function $_0F_1^{\gb/2}$, extending the $n=0$ result of \cite{Fo94b}.
\begin{prop}
Let $\gb a/2, \gb \in \Z_{\geq 0}$.  We have
\bea\label{32}
E^{\rm hard }_{\gb}(n;(0,s);\gb a/2) &=& A_\gb(n,a)s^{n+(\gb/2)n(n+a-1)}\e^{-\gb s/8} \nn \\
&& \times \int_0^1dy_1 \dots \int_0^1dy_n \prod_{j=1}^n (1-y_j)^{\gb a/2} \prod_{1\leq j<k\leq n}|y_k-y_j|^\gb \nn \\
&& \times\hspace{1mm} _0F_1^{(\gb/2)}({}_\_;a+2n;(s/4)^{\gb a/2},(sy_1/4)^\gb,\dots,(sy_n/4)^\gb),
\eea
where
\bea \label{32a}
A_\gb(n,a)={2^{-2n} \over n!} \Big ( {\beta \over 2} \Big )^n \bigg(\frac{\gb}{4}\bigg)^{n(a+n-1)\gb} \frac{(\Gamma(1+\gb/2))^n}{\prod_{j=0}^{2n-1} \Gamma(a\gb/2+1+j\gb/2)},
\eea
and in the argument of $_0F_1^{\gb/2}$ the notation $(u)^r$ means $u$ repeated $r$ times.  Furthermore, in the case $n=0$, this remains valid for general $\gb  > 0$.
\end{prop}
An integral representation of  $_0F_1^{(\gb/2)}$ allows for the rigorous determination of  the $x \rightarrow \infty$ asymptotic expansion of  $_0F_1^{(\gb/2)}
({}_\_;c+2(m-1)/\gb;(x)^m)$ for $c,2/\gb \in \Z^+$ \cite{Fo94b}, implying the corresponding asymptotic form of $E^{\mbox{\scriptsize{hard}}}_{\gb}(0;(0,s))$.
\begin{prop} (Forrester 1994)
Let $2/\gb \in \Z^+$, and $a\gb/2=m \in Z_{\geq0}$.  For $s \rightarrow \infty$ we have
\bea\label{r1}
E^{\rm hard }_{\gb}(0;(0,s);m)=\tau_{m,\gb}\bigg(  \frac{1}{s}\bigg)^{m(m+1)/2\gb-m/4}\e^{-\gb s/8+ms^{1/2}}\Bigg(1+O\bigg(\frac{1}{s^{1/2}}\bigg)\Bigg),
\eea
where
\bea\label{r2}
\tau_{m,\gb}^{\rm hard} =2^{(2/\gb-1)m}\bigg( \frac{1}{2\pi}\bigg)^{m/2}\prod_{j=1}^m \Gamma(2j/\gb).
\eea
\end{prop}
We see that (\ref{r1}) is in agreement with the log-gas prediction (\ref{CM1})  for general $a > -1, \gb>0$, and furthermore gives the explicit form of the constant in the asymptotic expansion (to use (\ref{r2}) for $m \notin \Z_{\geq 0}$ and check for example (\ref{3.26a}) requires an appropriate rewrite of the product using
(\ref{r}).

In the case $\gb=4$ of ${}_0F_1^{\gb/2}$, an integral representation not available for general $\gb$ shows that for  $s \rightarrow \infty$ and $y_1,\dots,y_n \approx 1$, \cite{Mu78}
\bea
{}_0F_1^{(\gb/2)}({}_\_;c;(s/4)^{\gb a/2},(sy_1/4)^\gb,\dots, (sy_n/4)^\gb) \hspace{6cm}\nn \\
 = {}_0F_1^{(\gb/2)} ({}_\_;c;(s/4)^{\gb(a+2n)/2} ) e^{\gb \sqrt{s}\sum_{j=1}^n(1-y_j)/2 }\Bigg( 1+O \bigg(\frac{1}{s^{1/2}}\bigg)\Bigg).
\eea
This, substituted in (\ref{32}) implies, as a conjecture, the extension of the asymptotic formula (\ref{31h}) to include the constant term.
\begin{conj}
For $\beta n \in \mathbb Z_{\ge 0}$, let
$$
\tau_{\beta a/2,\beta}^{\rm hard}(n) =
{2^{-(a+n)\beta n} \over n!}
\Big ( {\beta \over 2} \Big )^{n(a+n-1)\beta/2}
\prod_{j=1}^{\beta n} {\Gamma(a + 2j/\beta) \over (2 \pi)^{1/2}}
{\prod_{j=0}^{n-1} \Gamma(1 + (j+1)\beta/2) \over \prod_{j=n}^{2n-1} \Gamma(1 + (j+a)\beta/2)}.
$$
For $s \to \infty$ we have
\begin{equation}\label{Ef}
{E^{\rm hard }_{\gb}(n;(0,s);\gb a/2) \over E^{\rm hard }_{\gb}(0;(0,s);\gb a/2) } =
\tau_{\beta a/2,\beta}^{\rm hard}(n)
\exp \Big ( - \beta \Big \{  - \sqrt{s} n + \Big ( {n^2 \over 2} + {n a \over 2} \Big ) \log s^{1/2} \Big \} \Big )
\Big ( 1 +  O\Big ( {1 \over s^{1/2}} \Big ) \Big ).
\end{equation}
\end{conj}
We can check that (\ref{Ef})  is consistent with the results of Proposition \ref{P7}.

\subsection{Approach to unity of $E_\beta^{(\cdot)}(0;J)$ for $|J| \to 0$}
Generally the gap probability is given in terms of the $k$-point correlation functions $\{\rho_{(k)}^{(\cdot)}\}$ according to
$$
E_\beta^{(\cdot)}(0;J) = 1 - \int_J \rho_{(1)}^{(\cdot)}(x) \, dx + {1 \over 2!}  \int_J  \int_J   \rho_{(2)}^{(\cdot)}(x,y) \, dxdy - \cdots
$$
Thus the leading $|J| \to 0$ asymptotic form of $E_\beta^{(\cdot)}(0;J)$ is determined by the asymptotic form of 
$\rho_{(1)}^{\rm hard}(x)$ for $x \to 0$, $\rho_{(1)}^{\rm soft }(x)$ for $x \to \infty$ and $\rho_{(2)}^{\rm bulk}(x,y)$ for
$x,y \to 0$ (for $(\cdot)$ = bulk, $\rho_{(1)}(x) = 1$ and so gives no distinguishing information). The calculation
of the first and third is elementary \cite{Fo92j}, \cite{Fo94b}, while direct calculation of $\rho_{(1)}^{\rm soft}(x)$ is only
known for $\beta = 1$, and $\beta$ even \cite{DF06}. Collecting these together, we have the following result.

\begin{prop}
Let
$$
A_{a,\beta} = 4^{-(a+1)} (\beta/2)^{2a+1} {\Gamma(1 + \beta/2) \over \Gamma(1 + a) \Gamma(1 + a + \beta/2)}, \qquad
B_\beta = (\pi \beta)^\beta {(\Gamma(\beta/2 + 1))^3 \over \Gamma(\beta + 1) \Gamma(3 \beta/2 + 1)}.
$$
For $t \to 0$,
\begin{align}
E_\beta^{\rm hard}(0;(0,t);a) & = 1 - A_{a,\beta} \int_0^t s^a \, ds + O(t^{a+2}) \nn \\
E_\beta^{\rm bulk}(0;(0,t)) & = 1 - t + {1 \over 2} B_\beta \int_0^t \int_0^t (s_1 - s_2)^\beta \, ds_1 ds_2 + O(t^{\beta + 3}),
\end{align}
while for $t \to \infty$
\begin{equation}\label{tas}
E_\beta^{\rm soft}(0;(t,\infty))  = 1 - {\Gamma(1 + \beta/2) \over (4 \beta)^{\beta/2}} \int_t^\infty {e^{-2 \beta X^{3/2}/3} \over X^{3\beta/4 - 1/2}} \, dX +
O \Big ( \int_t^\infty {e^{-2 \beta X^{3/2}/3} \over X^{3\beta/4 + 1}} \, dX \Big ).
\end{equation}
\end{prop}

Two distinct derivations of (\ref{tas}) for general $\beta > 0$ are known, both involving use of a non-rigorous double scaling limit  \cite{Fo11, BN11}.
In \cite{DV11}, the stochastic differential equation characterization (recall  Section \ref{SDE}) is used to give a rigorous proof for general
$\beta > 0$, but without determining the prefactor of the integral.

\section{Other aspects}
\subsection{Numerical results}
Bornemann \cite{Bo08,Bo09} has given a detailed study of the numerical analysis relating to the precise numerical evaluation of spacing
distributions for $\beta = 1,2$ and 4, working from the Fredholm determinant forms. As an end product he has provided a suite of
Matlab programs implementing the theoretical procedures. The implementation in Matlab, with the arithmetic done in the hardware, means
that the tails of the spacing distributions cannot be computed: their numerical values written as decimals are typically smaller than $10^{-15}$,
and so double precision arithmetic typically truncates significant nonzero digits, leading to unreliable results. But with there being numerous
exact and conjectured results relating to the asymptotics of spacing distributions, there is much interest in implementing the theory of
\cite{Bo08,Bo09} using an arbitrary precision package. As a start, we have done this for the Fredholm determinant form for
$E_2^{\rm bulk}(0;(0,s))$ (in fact we have modified the procedure of \cite{Bo08,Bo09} by using instead of Gauss-Legendre or
Clenshaw-Curtis quadrature rules, the tanh-sinh quadrature rule (see e.g.~\cite{Ye06})). As a result we are able to tabulate
\begin{equation}\label{4.1}
r(s) = {E_2^{\rm b,as}(0;(0,s)) \over E_2^{\rm bulk}(0;(0,s))},
\end{equation}
where $E_2^{\rm b,as}(0;(0,s))$ is the asymptotic form of $E_2^{\rm bulk}(0;(0,s))$ as given by (\ref{3.23}), extended to the next two terms:
$1/(8 (\pi s)^2) + 5/(8 (\pi s)^4)$ (these follow from the Painlev\'e transcendent characterization of
$E_2^{\rm bulk}(0;(0,s))$  (see e.g.~\cite[\S 9.6.7]{Fo10})). The values in Table \ref{T1} clearly illustrate the accuracy of the
asymptotic expansion, even for relatively small values of $s$.

\begin{table}
\begin{center}
\begin{tabular}[t]{c||c}
$s$ & $r(s)$ \\\hline
 1 & 1.0046735914726577\\
 2 & 0.9998383226940526\\
 3 & 0.9999753765440204\\
 4 & 0.9999961026171116\\
 5 & 0.9999991096965057\\
 6 & 0.9999997235559452\\
 7 & 0.9999998946139279\\
 8 & 0.9999999537746553\\
 9 & 0.9999999775313906 \\
10 & 0.9999999881794448
\end{tabular}
\caption{\label{T1} Tabulation of the ratio of the asymptotic to exact bulk gap probability for $\beta = 2$.}
\end{center}
\end{table}

\subsection{Diluted spectra}
For a general one-dimensional point process, the generating function (\ref{dilute}) can also be interpreted as the probability that there are
no eigenvalues in the interval $J$, given that each eigenvalue has independently been deleted with probability $(1 - \xi)$. In this setting
the $|J| \to \infty$ asymptotics can readily be deduced, by making use of a heuristic analysis based on (\ref{n0})
\cite{BP04}.

\begin{conj}
For $0 < \xi < 1$ we have
\begin{equation}\label{e6}
E_\beta^{(\cdot)}(J;\xi) \mathop{\sim} \limits_{|J| \to \infty} e^{\langle n_J \rangle \log (1 - \xi)},
\end{equation}
where $\langle n_J \rangle$ is given by (\ref{n1}) for $(\cdot)=$ bulk, soft and hard.
\end{conj}

We see from (\ref{n1}) and (\ref{D1}), (\ref{CM1}), (\ref{CM1v}) that as a function of $s$ the decay exhibited by (\ref{e6}) is proportional to
the square root of the leading decay of $E_\beta^{(\cdot)}$. A method to prove (\ref{e6}) for $\beta = 2$, making use of (\ref{dilute1}),
has been given by Pastur and Shcherbina \cite{PS11}. Alternatively, for this $\beta$, (\ref{e6}) can be verified by using known asymptotics of the
Painlev\'e transcendent evaluations, as done for $(\cdot)=$ soft in \cite{BCP09}. 

An interesting feature of the asymptotic expansion of the 
relevant Painlev\'e transcendents with $0 < \xi < 1$ is that they contain oscillatory terms, in contrast to their asymptotic expansion with $\xi = 1$.
It is indeed the case that oscillations can clearly be seen in plots of ${d \over ds} E_2^{\rm soft}((s,\infty);\xi)$ with $0 < \xi < 1$ \cite{BCP09}.
Dyson \cite{Dy95} has combined Coulomb gas and Painlev\'e theory to deduce the asymptotic form $E_2^{\rm bulk}((0,s);\xi)$ when
$\xi \to 1$ and simultaneously $s \to \infty$, which is shown to involve an elliptic theta function; for fixed $\xi$ the asymptotic
expansion of the relevant Painlev\'e transcendent \cite{MT86} involves only trigonometric functions.

\subsection*{Acknowledgements}
The assistance of Tomasz Dutka for carrying out the numerical work of Section 4.1 during the 2012 Vacation Scholarship program in the
Department of Mathematics and Statistics at the University of Melbourne, and the assistance of Mark Sorrell in the preparation of
the manuscript, is acknowledged. 
Thanks are due to the organizers of the MSRI program `Random matrices, interacting particle systems and integrable systems' for providing
financial support and a stimulating environment. This research has been supported by the  Australian Research Council.

\providecommand{\bysame}{\leavevmode\hbox to3em{\hrulefill}\thinspace}
\providecommand{\MR}{\relax\ifhmode\unskip\space\fi MR }
% \MRhref is called by the amsart/book/proc definition of \MR.
\providecommand{\MRhref}[2]{%
  \href{http://www.ams.org/mathscinet-getitem?mr=#1}{#2}
}
\providecommand{\href}[2]{#2}


\begin{thebibliography}{10}

\bibitem{BBD07}
J.~Baik, R.~Buckingham, and J.~DiFranco, \emph{Asymptotics of {T}racy-{W}idom
  distributions and the total integral of a {P}ainlev\'e {II} function},
  Commun. Math. Phys. \textbf{280} (2008), 463--497.
  
  \bibitem{Ba04}
E.W.~Barnes, \emph{The theory of the multiple gamma function}, Trans.~Cam.~Phil.~Soc.
\textbf{19} (1904), 374--425.


\bibitem{BTW92}
E.L.~Basor and C.A.~Tracy and H.~Widom, \emph{Asymptotics of level spacing distributions for
random matrices}, Phys. Rev. Lett. \textbf{69} (1992), 5--8.

 \bibitem{BP04} O.~Bohigas and M.P.~Pato, \emph{Missing levels in correlated spectra},
  Phys.~Lett.~B \textbf{595} (2004), 171--176.
  
 \bibitem{BCP09} O.~Bohigas, J.X.~de Carvalho and M.P.~Pato, \emph{Deformations of the
    Tracy-Widom distribution}, Phy. Rev. E \textbf{79} (2009), 031117 [6 pages]

\bibitem{Bo09}
F.~Bornemann, \emph{On the numerical evaluation of distributions in random
  matrix theory: a review with an invitation to experimental mathematics},
  Markov Processes Relat. Fields \textbf{16} (2010), 803--866.

\bibitem{Bo08}
\bysame, \emph{On the numerical evaluation of {F}redholm determinants}, Math.
  Comp. \textbf{79} (2010), 871--915.

\bibitem{BEMN10}
G.~Borot, B.~Eynard, S.N. Majumdar, and C.~Nadal, \emph{Large deviations of the
  maximal eigenvalue of random matrices}, J. Stat. Mech. \textbf{2011} (2011),
  P11024.

\bibitem{BN11}
G.~Borot and C.~Nadal, \emph{Right tail expansion of {T}racy-{W}idom beta
  laws}, arXiv:1111.2761, 2011.
  
  \bibitem{BMS11}
A.~Brini and M.~Marino and S.~Stevan, \emph{The uses of the refined matrix model recursion},
J.~Math.~Phys. {\bf 52}, (2011) 052305(24 pages)   

\bibitem{CM94}
Y.~Chen and S.M. Manning, \emph{Asymptotic level spacing of the laguerre
  ensemble: a {C}oulomb fluid approach}, J. Phys. A \textbf{27} (1994),
  3615--3620.

\bibitem{CM96}
\bysame, \emph{Some eigenvalue distribution functions of the laguerre
  ensemble}, J. Phys. A \textbf{29} (1996), 7561--7579.
  
\bibitem{CL95}
O.~Costin and J.L.~Lebowitz, \emph{Gaussian fluctuations in random matrices}, Phys.~Rev.~Lett.
\textbf{75} (1995), 69--72

\bibitem{DM06}
D.S. Dean and S.N. Majumdar, \emph{Large deviations of extreme eigenvalues of
  {G}aussian random matrices}, Phys. Rev. Lett. \textbf{97} (2006), 160201.

\bibitem{DM08}
\bysame, \emph{Extreme value statistics of eigenvalues of {G}aussian random
  matrices}, Phys. Rev. E \textbf{77} (2008), 041108.

\bibitem{DIK08}
P.~Deift, A.~Its, and I.~Krasovsky, \emph{Asymptotics of the {A}iry kernel
  determinant}, Commun. Math. Phys. \textbf{278} (2008), 643--678.

\bibitem{DKV11}
P.~Deift, A.~Its, and J.~Vasilevska, \emph{Asymptotics for a determinant with a
  confluent hypergeometric kernel}, Int. Math. Res. Not. \textbf{2011} (2011),
  2117--2160.

\bibitem{DIZ96}
P.A. Deift, A.R. Its, and X.~Zhou, \emph{A {Riemann-Hilbert} approach to
  asymptotic problems arising in the theory of random matrices and also in the
  theory of integrable statistical mechanics}, Ann. Math. \textbf{146} (1997),
  149--235.

\bibitem{DF06}
P.~Desrosiers and P.J. Forrester, \emph{Hermite and {L}aguerre
  $\beta$-ensembles: asymptotic corrections to the eigenvalue density}, Nucl.
  Phys. B \textbf{743} (2006), 307--332.

\bibitem{DV11}
L.~Dumaz and B.~Vir\'ag, \emph{The right tail exponent of the
  {T}racy-{W}idom-beta distribution}, arXiv:1102.4818, 2011.

\bibitem{DE02}
I.~Dumitriu and A.~Edelman, \emph{Matrix models for beta ensembles}, J. Math.
  Phys. \textbf{43} (2002), 5830--5847.

\bibitem{Dy62e}
F.J. Dyson, \emph{Statistical theory of energy levels of complex systems {II}},
  J. Math. Phys. \textbf{3} (1962), 157--165.

\bibitem{Dy76}
\bysame, \emph{Fredholm determinants and inverse scattering problems}, Commun.
  Math. Phys. \textbf{47} (1976), 171--183.

\bibitem{Dy95}
\bysame, \emph{The {C}oulomb fluid and the fifth {Painlev\'e} transcendent},
  Chen Ning Yang (S.-T. Yau, ed.), International Press, Cambridge MA, 1995,
  p.~131.

\bibitem{Eh06}
T.~Ehrhardt, \emph{Dyson's constant in the asymptotics of the {F}redholm
  determinant of the sine kernel}, Commun. Math. Phys. \textbf{262} (2006),
  317--341.

\bibitem{Eh10}
\bysame, \emph{The asymptotics of a {B}essel-kernel determinant which arises in
  {R}andom {M}atrix {T}heory}, Advances Math. \textbf{225} (2010), 3088--3133.

\bibitem{FS95}
M.~Fogler and B.I. Shklovskii, \emph{The probability of an eigenvalue
  fluctuation in an interval of a random matrix spectrum}, Phys. Rev. Lett.
  \textbf{74} (1995), 3312--3315.

\bibitem{Fo92j}
P.J. Forrester, \emph{Selberg correlation integrals and the $1/r^2$ quantum
  many body system}, Nucl. Phys. B \textbf{388} (1992), 671--699.

\bibitem{Fo93a}
\bysame, \emph{The spectrum edge of random matrix ensembles}, Nucl. Phys. B
  \textbf{402} (1993), 709--728.

\bibitem{Fo94b}
\bysame, \emph{Exact results and universal asymptotics in the {Laguerre} random
  matrix ensemble}, J. Math. Phys. \textbf{35} (1994), 2539--2551.

\bibitem{Fo06c}
\bysame, \emph{Hard and soft edge spacing distributions for random matrix
  ensembles with orthogonal and symplectic symmetry}, Nonlinearity \textbf{19}
  (2006), 2989--3002.

\bibitem{Fo09}
\bysame, \emph{A random matrix decimation procedure relating $\beta = 2/(r+1)$
  to $\beta = 2(r+1)$}, Commun. Math. Phys. \textbf{285} (2009), 653--672.
  
\bibitem{Fo10}
\bysame, \emph{Log-gases and random matrices}, Princeton University Press,
  Princeton, NJ, 2010.

\bibitem{Fo11}
\bysame, \emph{Spectral density asymptotics for {G}aussian and {L}aguerre
  $\beta$-ensembles in the exponentially small region}, J. Phys. A \textbf{45} (2012), 075206
  
  \bibitem{FR01}
P.J. Forrester and E.M. Rains, \emph{Inter-relationships between orthogonal, unitary and symplectic matrix
ensembles}, In P.M. Bleher and A.R. Its, editors,
\emph{Random matrix models and their applications}. volume 40 of \emph{Mathematical Sciences 
Research Institute Publications}, pages 171-208. Cambridge University Press, United Kingdom, 2001.


  
\bibitem{FW01a}
P.J. Forrester and N.S. Witte,  \emph{Application of the $\tau$-function theory of {Painlev\'e}
equations to random matrices: {PV}, {PIII}, the {LUE}, {JUE} and
{CUE}}, Commun. Pure Appl. Math. \textbf{55} (2002), 679--727.



\bibitem{FW11}
\bysame, \emph{Asymptotic forms for hard and soft edge
  general $\beta$ conditional gap probabilities}, Nucl. Phys. B \textbf{859} (2012), 321--340.

\bibitem{Fu64}
W.H.J. Fuchs, \emph{On the eigenvalues of an integral equation arising in the
  theory of band-limited signals}, J. Math. Anal. Appl. \textbf{9} (1964),
  317--330.

\bibitem{Ga61}
M.~Gaudin, \emph{Sur la loi limite de l'espacement des valeurs propres d'une
  matrice al\'eatoire}, Nucl. Phys. \textbf{25} (1961), 447--458.
  
  \bibitem{KO98}
  K.~Katayama and M.~Ohtsuki, \emph{On the multiple gamma-functions},
  Tokyo J. Math. \textbf{21} (1998), 159--182.
  
  \bibitem{Ki08} R.~Killip, \emph{Gaussian fluctuations for $\beta$ ensemble}, Int. Math. Res. Notices
  \textbf{2008} (2008), rnn007

\bibitem{KS06}
R.~Killip and M.~Stoiciu, \emph{Eigenvalue statistics for {CMV} matrices: from
  {P}oisson to clock via circular beta ensembles}, Duke Math. J. \textbf{146}
  (2009), 361--399.

\bibitem{Kr04}
I.~Krasovsky, \emph{Gap probability in the spectrum of random matrices and
  asymptotics of polynomials orthogonal on the unit circle}, Int. Math. Res.
  Not. \textbf{2004} (2004), 1249--1272.

\bibitem{Me91}
M.L. Mehta, \emph{Random matrices}, 2nd ed., Academic Press, New York, 1991.

\bibitem{JMC72}
M.L. Mehta and J.~des Cloizeaux, \emph{The probabilities for several
  consecutive eigenvalues of a random matrix}, Indian J. Pure Appl. Math.
  \textbf{3} (1972), 329--351.
  
  \bibitem{MT86}
B.M. McCoy and S. Tang, \emph{Connection formulae for {Painlev\'e} {V} functions},
Physica D, \textbf{19} (1986), 42--72   

\bibitem{Mu78}
R.J. Muirhead, \emph{Latent roots and matrix variates: a review of some
  asymptotic results}, Ann. Stat. \textbf{6} (1978), 5--33.
  
  \bibitem{PS11}
L.~Pastur and M.~Shcherbina, \emph{Eigenvalue distribution of large random matrices}, American Mathematical Society,
  Providence, RI, 2011.

\bibitem{Po65}
C.E. Porter, \emph{Statistical theories of spectra: fluctuations}, Academic
  Press, New York, 1965.
  
  \bibitem{QHS93}
  J.R.~Quine and S.H.~Heydari and R.Y.~Song, \emph{Zeta regularized products},
  Trans. Am. Math. Soc. \textbf{338}, 213--231.

\bibitem{RR08}
J.~Ramirez and B.~Rider, \emph{Diffusion at the random matrix hard edge},
  Commun. Math. Phys. \textbf{288} (2009), 887--906.

\bibitem{RRV06}
J.~Ramirez, B.~Rider, and B.~Virag, \emph{Beta ensembles, stochastic {A}iry
  spectrum, and a diffusion}, J. Amer. Math. Soc. \textbf{2011} (2011),
  919--944.

\bibitem{RRZ11}
J.A. Ramirez, B.~Rider, and O.~Zeitouni, \emph{Hard edge tail asymptotics},
  Elec. Comm. in Probab. \textbf{16} (2011), 741--752.
  
  
  \bibitem{Sh80}
  T.~Shintani, \emph{A proof of the classical {K}ronecker limit formula}, 
  Tokyo J. Math. \textbf{3} (1980), 191--199.
  
  \bibitem{So00}
  A.~Soshnikov, \emph{Determinantal random point fields}, Russian Math. Surveys \textbf{55} (2000), 923--975.

\bibitem{TW93}
C.A. Tracy and H.~Widom, \emph{Introduction to random matrices}, Geometric and
  quantum aspects of integrable systems (G.F. Helminck, ed.), Lecture Notes in
  Physics, vol. 424, Springer, New York, 1993, pp.~407--424.

\bibitem{TW94a}
\bysame, \emph{Level-spacing distributions and the {Airy} kernel}, Commun.
  Math. Phys. \textbf{159} (1994), 151--174.

\bibitem{TW94b}
\bysame, \emph{Level-spacing distributions and the {Bessel} kernel}, Commun.
  Math. Phys. \textbf{161} (1994), 289--309.

\bibitem{VV07}
B.~Valk\'o and B.~Vir\'ag, \emph{Continuum limits of random matrices and the
  {B}rownian carousel}, Inv. Math. \textbf{177} (2008), 463--508.

\bibitem{VV08}
\bysame, \emph{Large gaps between random eigenvalues}, Ann. Prob. \textbf{38}
  (2008), 1263--1279.

\bibitem{Ye06}
L. Ye, \emph{Numerical quadrature: theory and computation}, MSc. thesis,
  Dalhousie University, 2006.

\end{thebibliography}
\end{document}